\documentclass[useAMS,usenatbib]{mn2e}
\usepackage{graphicx}

\def\simlt{$\; \buildrel < \over \sim \;$}
\def\simlt{\lower.5ex\hbox{\simlt}}
\def\gtsima{$\; \buildrel > \over \sim \;$}
\def\simgt{\lower.5ex\hbox{\gtsima}}
\def\fe{[Fe/H]}

\title[The Canis Major galaxy]
{Detection of the Canis Major galaxy at
($l;b$)=(244$\degr$;-8$\degr$) and in the
background of Galactic open clusters\thanks{Based on archive data from 
observations made with the ESO-MPI WFI telescope, at ESO, La Silla, Chile 
under programme 170.A-0789(A)}.}
\author[M. Bellazzini et al.]{
M. Bellazzini$^{1}$\thanks{E-mail: bellazzini@bo.astro.it}, R. Ibata$^{2}$,
L. Monaco$^{1,5}$, N. Martin$^{2}$, M.J. Irwin$^{3}$, G.F. Lewis$^{4}$\\
$^{1}$
INAF - Osservatorio Astronomico di Bologna, via Ranzani 1, 40127, Bologna,
Italy\\
$^{2}$
Observatoire de Strasbourg, 11, rue de l'Universit\'e, F-67000, Strasbourg, 
France\\
$^{3}$
Institute of Astronomy, Madingley Road, Cambridge, CB3 0HA, U.K.\\
$^{4}$
Institute of Astronomy, School of Physics, A29, University of Sydney, NSW
2006, Australia\\
$^{5}$
INAF - Osservatorio Astronomico di Trieste, via Tiepolo 11, 34131, Trieste,
Italy}
\date{5 August 2004, Accepted for publication by MNRAS.}

\begin{document}

\pagerange{\pageref{firstpage}--\pageref{lastpage}} \pubyear{2003}

\maketitle

\label{firstpage}

\begin{abstract}
We report on the detection of Main Sequence stars belonging to the
recently identified Canis Major galaxy in a field located at $\simeq 4.2\degr$ 
from the center of the stellar system.
With Main Sequence fitting we obtain a distance modulus
$(m-M)_0=14.5\pm 0.3$ to the dwarf, corresponding to a distance of 
$D_{\sun}\simeq 8.0 \pm 1.2$ kpc, in full agreement with previous estimates
based on the photometric parallax of M-giants. From the comparison with
theoretical isochrones we constrain the age of the main population of the Canis
Major system in the range $\sim 4-10$ Gyr. A blue plume of likely younger
stars (age $< 1-2$ Gyr) is also identified.
The available Colour Magnitude Diagrams of open clusters that may be projected
onto the main body of Canis Major are also briefly analyzed. 
The position, distance and stellar population of the old open clusters AM-2 and
Tombaugh~2 strongly suggest that they are physically associated with the Canis
Major galaxy. Using our own photometry and data from 2MASS and the GSC2.2
surveys we demonstrate that the claim by Momany et al. that the CMa overdensity
is entirely due to the Galactic warp is not supported by the existing
observations, once all the available pieces of information are taken into
account. It is shown that the CMa overdensity clearly emerges at a heliocentric
distance of $\sim 8$ kpc above any overdensity possibly produced by the Galactic
warp.
\end{abstract}


\section{Introduction}

The process of hierarchical merging \citep{wr78,wf91} is generally accepted as
the driving mechanism of the formation of giant galaxies. The study of the local
(e.g. Galactic or in the Local Group) relics of such processes provides an
unprecedentedly detailed insight of the current cosmological model as well as
a formidable testbed for theories of galaxy formation. The case of the
Sagittarius dwarf galaxy \cite[Sgr dSph]{s1,s2} is emblematic, since we
are presently observing it during its disruption into a giant stream
contributing to the continued assembly of the Galactic halo with stars 
\cite[see][and references therein]{ibata01,ibata02,ivez,newberg,majewski} and 
globular clusters \citep{bell03a,bell03b,bell03c}. A similar case has been 
reported in M31 \citep{m31s1,m31s2} and systematic searches in galaxies outside
the Local Group are beginning \citep{poh}.

An even more interesting case is provided by the recently discovered ring
structure \cite[the Monoceros Ring, hereafter the Ring, for
brevity][]{newberg,yanny,ibaring,majewski,rochap} near the Galactic plane. 
The more recent observational results \citep{crane,martin} suggests that
the Ring is the stream remnant of an in-plane accretion that may be contributing
to the build-up of the thick disk \cite[see also][]{helmi,abadi}. 

In particular
\citet[][hereafter Pap-I]{martin}, while tracing the whole extent of the Ring
using M-giants from the 2MASS database\footnote{See Cutri et al. (2003), 
Explanatory Supplement to the 2MASS All Sky Data Release,  
http://www.ipac.caltech.edu/2mass/releases/allsky/doc/explsup.html} discovered
a large overdensity of "Ring-like" M-giants in the Canis Major constellation.
The elliptical shape, the overall structure of this overdensity and
its coincidence with a noticeable grouping of globular clusters
\citep{bell03c} led
\citet{martin} to the conclusion that it is the relic of the dwarf galaxy whose
disruption generated the Ring. The discovered relic (hereafter Canis Major
galaxy, CMa) is approximately centered at galactic coordinates
(l,b)$\sim(240^o,-7^o)$, it has a FWHM extension in the sky of $\sim 12 \degr$
in the latitude direction and it is located at $\sim 7$ kpc from the Sun 
(i.e., $(m-M)_0\sim 14.25$), as obtained from the photometric parallax to 
M-giants introduced by \citet{majewski}. The mass, luminosity and characteristic
dimensions of Canis Major appear quite similar to those of the Sgr dSph. The
similarity with Sgr extends to other relevant characteristics: the two objects
seem to host a similar grouping of globular clusters, and the upper RGB of both
galaxies is dominated by stars of similar colour (M-giants), which are
present in similar numbers in both galaxies. The analogy between
the two systems may provide a useful guideline in the study of the newly
discovered galaxy. 

\citet{martin} obtained a near infrared (NIR)
Colour Magnitude Diagram (CMD) of the discovered structure (e.g., their Fig.~8)
showing a Red Giant Branch (RGB) extending to $(J-K_S)_0\simeq 1.3$ and a
pronounced Red Clump (RC) at $K_{S0}\ge 12.5$ and  $(J-K_S)_0\simeq 0.6$. A Blue
Plume (BP) of stars (of more uncertain nature) is also detected around 
$(J-K_S)_0\simeq 0.3$. The Main Sequence (MS) of the CMa population is beyond
the reach of 2MASS photometry. The degeneracy among distance, age, metallicity
and reddening (which is particularly high and highly variable at such low
latitudes) prevents any firm conclusion about the nature of the stellar
population in the CMa structure. Furthermore the distance scale based on
photometric parallax of M-giants is intrinsically uncertain 
\citep{bell03b,majewski} and it is likely affected by a systematic effect 
leading to the underestimate of true distances \citep{martin,newberg2}. 
Hence, the
detection of fainter stars belonging to the newly discovered galaxy, the access
to alternative distance indicators than M-giants and the study of its stellar
population in other wavelength ranges (than the NIR) are highly desirable
to confirm the discovery presented in Pap-I and to obtain a clearer
characterization of the system.

Here we present the first detection of the Main Sequence (MS) of the CMa 
system in a
wide field located $\sim 4.2\degr$ apart from its center, from deep B 
and V photometry of very recent ESO-WFI images taken from the ESO archive.
We also take advantage of the fact that the low latitude
region of sky covered by the CMa galaxy hosts a number of galactic open
clusters (OC), most of them in the foreground. 
Thus we searched for existing wide
field CCD photometry of these open clusters to look for evidence of a
background population that may be related to the CMa galaxy. 
Unfortunately, modern wide field photometries of open clusters are quite rare
since, typically, a field of few arcmin is sufficient to characterize the
population of these clusters and the background population is an undesired
contaminant for these kind of studies. Nevertheless,
some interesting clues of the presence of CMa stars in these fields
have been obtained, providing useful indications on the
spatial extent of the system and guidance for future observations.

The list of the fields considered in the current study is reported in Table~1,
where we report the position of the fields in Galactic coordinates, the field of
view covered by the available observations, the angular distance from the center
of CMa, an estimate of the reddening and the source of the data. 
The position of the various fields with respect to the CMa
system is sketched in Fig.~1.
Most of the photometry (catalogues in electronic form) used in this paper have
been retrieved from the WEBDA\footnote{\tt
http://obswww.unige.ch/webda/webda.html} database \citep{mermi}. The adopted
reddening values are taken from the maps by \citet[][SFD98]{schlegel}, if not
otherwise stated. All over the paper we adopt the reddening laws by \citet{dean}
for the (Cousins') I band and by \citet{rl85} for all the other passbands.

We anticipate that the data presented in this paper are not
sufficient to definitely break the above quoted {\em
distance/age/metallicity/reddening} degeneracy and that reliable spectroscopic
abundances and dedicated optical photometry are still needed to obtain a
fully consistent picture of the main characteristics of the galaxy. On the other
hand, the results presented here should be regarded as a significant progress 
in the understanding of the CMa system as well as a confirmation of its 
existence.

Criticizing the results presented in \citet{martin} and in a previous
preliminary version of the present paper circulated as a preprint, 
\citet[][hereafter M04]{moma4}
claimed that the CMa overdensity is completely due to the North-South asymmetry
produced by the Galactic warp \cite[see][and references therein]{yusi,kuij}.
We discuss this point in Sect.~4, demonstrating that the M04 hypothesis is
clearly confuted by several independent observational evidences.

\begin{table*}
 \centering
 \begin{minipage}{140mm}
  \caption{Explored Fields.}
  \begin{tabular}{@{}lrrcccc@{}}
  \hline
   Field Name & l$\degr$ & b$\degr$ & f.o.v. & dist\footnote{Angular distance from the centroid of
   CMa, in degrees.}&E(B-V)\footnote{From the reddening maps by
   \citet{schlegel}} & Ref. \\
  \hline
F-XMM   & 244.2 & -8.2 & $34\arcmin\times 33\arcmin$ &4.2$\degr$ &0.14 & This work\\
NGC~2477 (B,V) & 253.6 & -5.8 & $34\arcmin\times 33\arcmin$ & 13.6$\degr$&0.77 & \citet{moma}
\footnote{The (V,I) photometry by \citet{kas} has been also considered
(f.o.v=$14.7\arcmin\times 14.7\arcmin$).}\\
Tombaugh~1& 232.3 & -7.3 & $7.0\arcmin\times 7.0\arcmin$ & 7.7$\degr$&0.43& \citet{tom1}\\
Berkeley~33& 225.4 & -4.6 & $10.5\arcmin\times 10.5\arcmin$ & 14.8$\degr$&0.75& \citet{be33}\\
NGC~2204   & 226.0 & -16.1& $14.7\arcmin\times 14.7\arcmin$& 15.9$\degr$&0.10&\citet{kas} \\
Melotte 66 & 259.6 & -14.3& $14.7\arcmin\times 14.7\arcmin$& 20.2$\degr$&0.22&\citet{kas} \\
NGC~2243   & 239.5 & -18.0& $15.0\arcmin\times 15.0\arcmin$& 10.5$\degr$&0.07 & \citet{2243kal}\\
&{\bf Open Clusters}\footnote{Possibly associated with CMa, see Sect.~5.}&  & & & & \\
Arp-Madore 2& 248.1&-5.9 & $5.5\arcmin\times 5.5\arcmin$& 8.3$\degr$ &0.73&\citet{am2} \\
Tombaugh~2  & 232.9& -6.9& $10.4\arcmin\times 10.4\arcmin$& 6.9$\degr$&0.39 & \citet{tom2}\\
\hline
\end{tabular}
\end{minipage}
\end{table*}


\begin{figure}
\includegraphics[width=84mm]{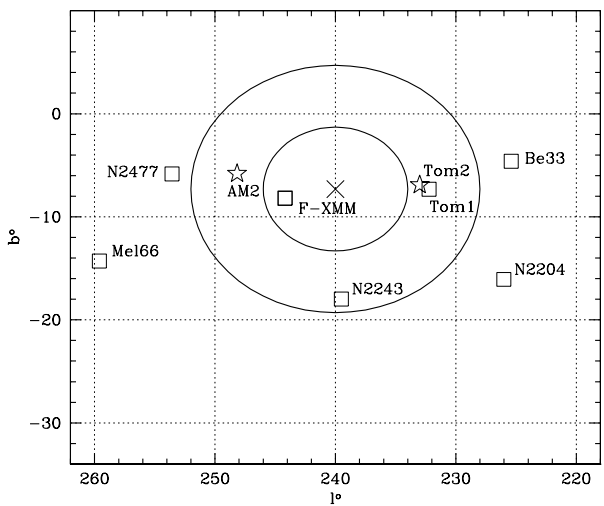} 
\caption{Relative positions on the sky (Galactic coordinates) of the CMa galaxy
and of the fields considered in the present analysis. The center of CMa is
represented as a cross, the two concentric circles have radius r$=6^o$ deg and 
r$=12^o$, corresponding to the HWHM and FWHM of the distribution of CMa 
M-giants as derived in Pap-I, respectively. The clusters that are possible
members of the CMa system are represented as open stars.
}
\end{figure}

\begin{figure*}
\includegraphics[width=168mm]{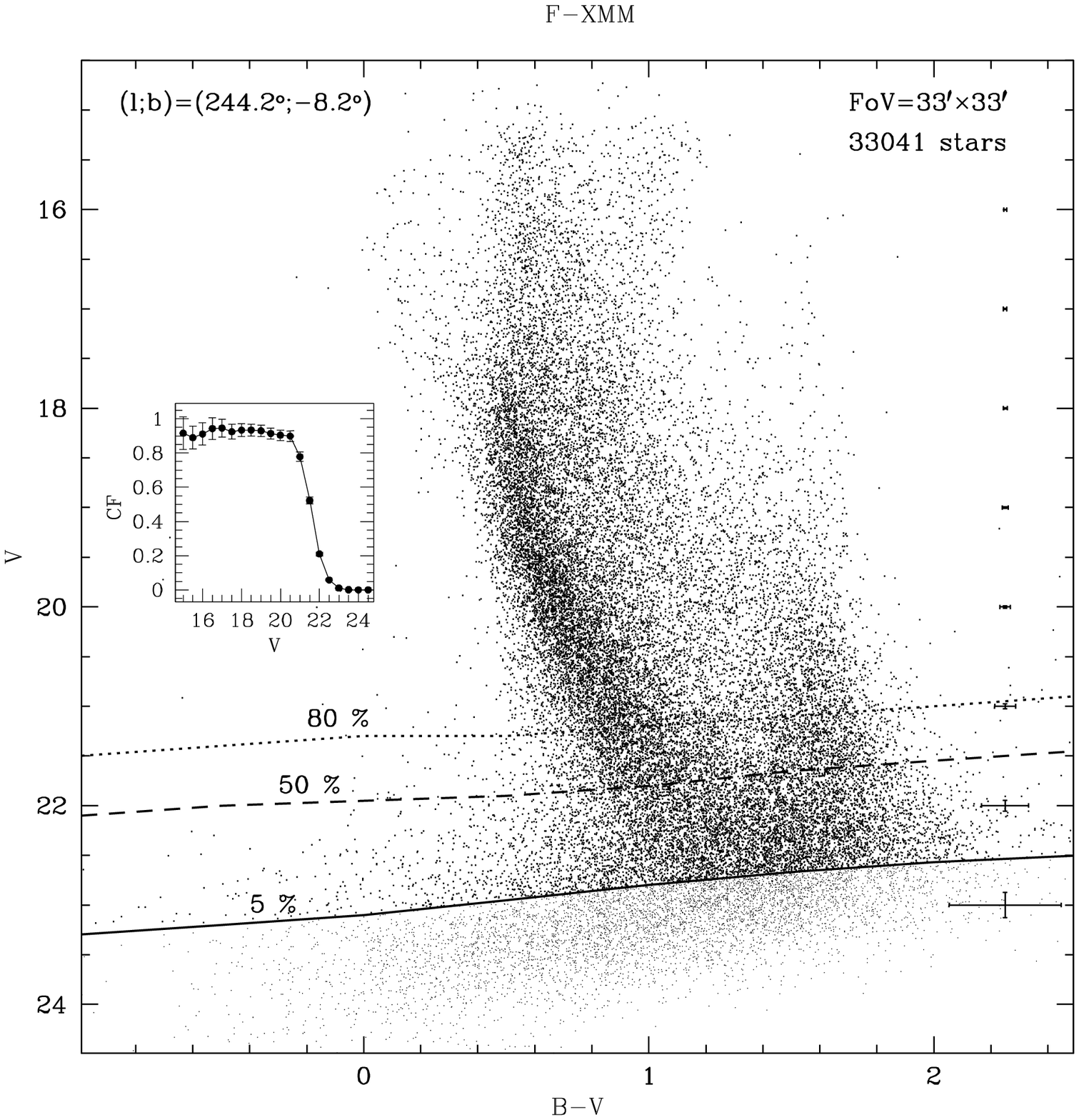} 
\caption{CMD of the $\simeq 0.5\degr \times 0.5\degr$ XMM field, located at
$\simeq 4.2\degr$ from the center of the CMa system. 
The error bars at $B-V\simeq 2.2$
are the average photometric errors as a function of magnitude.
The dotted, dashed and continuous
lines indicate the locus of the CMD where the joint 
Completeness Factor (CF) is 80\%, 50\% and
5\%, respectively. The joint CF as a function of
V magnitude (computed at $B-V=1.0$) is shown in the inset. 
}
\end{figure*}

\section{The MS of the Canis Major galaxy at $(l;b)=(244.2\degr;-8.2\degr)$.}

\subsection{Observational material and data reduction}

We retrieved from the ESO archive a set of deep B and V images of a field 
located
at (l;b)=($244.2\degr;-8.2$\degr), at $\simeq 4.2\degr$ from the center of CMa. 
The images were taken with the WFI camera\footnote{The WFI camera is a mosaic of 
eight
$2048\times 4096$ pixels chips, with pixel scale $0.238$ arsec/px and with a
total field of view of $\simeq 34\arcmin\times 33\arcmin$. See 
http://www.ls.eso.org/lasilla/Telescopes/2p2T/E2p2M/WFI/ for further details.} 
mounted at the 2.2m telescope at ESO (La Silla,
Chile) as part of the large programme {\em Public Imaging Survey: WFI Follow-up of 
XMM-Newton Serendipitous Fields} (170.A-0789(A)). We will refer to this
field  as the XMM Field (F-XMM), following the name of the
original programme. All the calibrating images (bias, flat-fields, standard star fields)
related to the same observational run were also retrieved.

To obtain stellar photometry of this field we selected the
best-seeing, lowest airmass ($AM<1.1$), V and B images of the 
set taken during a photometric night in which
observations of \citet{land} standard fields, taken with the same instrument, are 
also available (the night of February 1, 2003). 
The seeing and the exposure time are $0.88\arcsec$ FWHM and 
440 s for the V image, and $1.07\arcsec$ FWHM and 360 s for the B image,
respectively. The images were corrected for bias and flat-field using standard
procedures within the IRAF environment {\em mscred} and were reduced using
the PSF-fitting software DoPHOT \citep{doph}. The images were searched for
sources adopting a $3\sigma$ threshold above the background and modeling the
spatial variation of the PSF with a quadratic polynomial 
\cite[essentially the same approach adopted in][and described in more detail in
those papers]{umi,n5634}. Only sources classified as bona-fide stars by DoPHOT
were retained in the final catalogue of instrumental magnitudes, that was 
obtained by cross-correlating the individual B and V catalogues. All the sources
having a photometric error (in B or in V) larger than 3 times the average error
at their magnitude were also excluded from the catalogue. The instrumental
photometry has been corrected for the geometric distortion affecting the WFI
camera \cite[due to the effect of scattered light, see][]{manfroid} with a
simple empirical formula derived by M. Irwin from the photometry of several
Landolt's standard fields, that provides corrections very similar to the 
complex model provided by ESO \footnote{See
www.ls.eso.org/lasilla/Telescopes/2p2T/E2p2M/WFI/zeropoints/}.
The degree of crowding of the XMM Field is quite low, hence it has been easy to
select a large number ($>50$) of bright and isolated stars to obtain robust 
estimates of aperture corrections. 

The instrumental magnitudes have been transformed to the Johnson-Cousins system
with the calibrating equations

\begin{equation}
V-v=-0.134(b-v)+23.97 ~~(r.m.s.= 0.017 ~mag)
\end{equation}

\begin{equation}
B-b=0.337(b-v)+24.77 ~~(r.m.s.= 0.023 ~mag)
\end{equation}

\noindent
(where capital letters indicate calibrated magnitudes and lower case letters
indicate instrumental magnitudes), obtained from observations of Landolt's
field SA98. The extended list of standard stars by \citet{stet} has been
adopted. The atmospheric extinction coefficients estimated on
April 13, 2003 have been taken from the WFI homepage.

\subsection{Artificial star experiments}

To characterize the completeness of the F-XMM sample described above we 
performed
a set of artificial star experiments. The artificial stars were extracted from
a Luminosity Function (LF) similar to the observed one, modeled with the same
PSF model used for the photometry (including spatial variations), and added onto 
the original frames ($\sim 2000$ at each run and adopting the strategy
described in \citet{lf288,umi}, to avoid interferences between artificial stars) 
that were
then reduced in the same way as the original scientific images. A total of
$\simeq 30000$ stars were simulated in the V and B images. The completeness
factor $CF=N_{recovered}/N_{simulated}$ as a function of magnitude was
determined separately for each passband, the joint completeness factor at any
given colour is then obtained by the product $CF=CF_V\times CF_B$.

\begin{figure*}
\includegraphics[width=168mm]{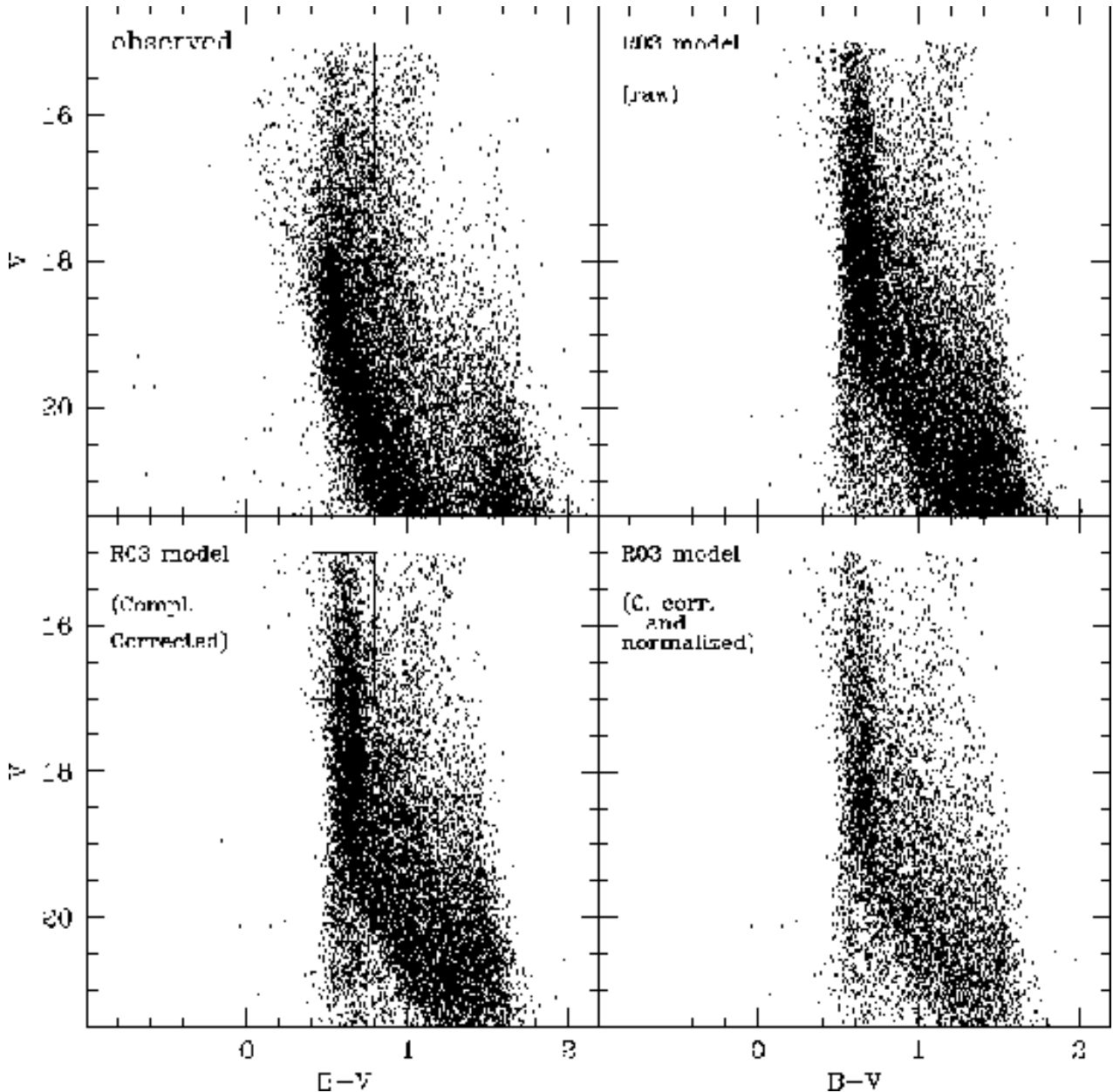} 
\caption{Comparison with the observed CMD of F-XMM (upper left panel)
and the synthetic CMD from the Galactic model by \citet{r03}.
Upper right panel: raw synthetic CMD; lower left panel: synthetic CMD corrected
for incompleteness; lower right panel: synthetic CMD corrected for 
incompleteness and normalized to have the same number of stars as the observed
CMD in the selection box plotted in CMDs on the left-hand panels.}
\end{figure*}

\subsection{The Colour Magnitude Diagram}

The CMD of the XMM Field is shown in Fig.~2. 
The dotted, dashed and continuous
lines indicate the locus of the CMD where the joint CF is 80\%, 50\% and
5\%, respectively. In the following analysis we will ignore the sources
fainter than the CF=5\% limit.

Even a first glance to Fig.~2 shows that the F-XMM CMD is dominated by an
unexpected sequence running from $(V;B-V)\sim(18.0;0.6)$ to 
$(V;B-V)\sim(23.0;1.3)$, superposed on the typical field population. 
The morphology of the feature is typical of the MS
of an intermediate-old stellar population. The possible Turn Off point (TO) of
this sequence, at $V\sim 19$ is slightly bluer than the blue edge of the
vertical plume of TO field stars located at various distances along the line of
sight \citep{gw87,unavane,prandoni,morrison}, 
that is visible for $V<18$ around $B-V\simeq 0.45$ in this diagram. Note that
the same characteristic (i.e. TO stars bluer than the average field TO  stars)
was one of the key elements that allowed the discovery of the Ring structure,
that was first identified as an excess of stars of spectral type A-F 
\citep{newberg,yanny}. The sequence (as well as the overall appearance of the
CMD) is also very similar to the Ring MS found by \citet{ibaring}.
As in the CMDs presented by these authors, the Sub Giant Branch (SGB) and the 
base of the Red Giant Branch (RGB) of the unexpected population are hidden 
by the contamination of Galactic field stars.
The saturation level of the photometry, occurring at $V\simeq 15.0$, 
prevents the detection of the Red Clump (RC) of Helium burning stars associated
with the population, that are likely brighter than this limit.

Other noticeable features of the CMD are (1) a sparse blue plume at $B-V\sol
0.4$ and $V\sol 18.5$, and (2) the vertical wedge-shaped plume of the local M
dwarfs around $B-V\simeq 1.6$.

\subsection{Comparison with a Milky Way model}

Lacking a suitable Control Field to compare with F-XMM we are forced to rely
on a synthetic model of the stellar populations of the Milky Way. We adopt, here
and in the following, the Galactic model by 
\citet[][hereafter R03]{r03}, from which it is possible to obtain synthetic CMDs
for a given direction and field of view through a web interface. 
The model includes all the known
Galactic components (thin and thick disc, halo and bulge) and takes into account
the effects of the disc warp and flare. The last characteristic is particularly
useful in the present context, since the (stellar) Galactic warp reaches its 
maximum projected density at $l\simeq 270\degr$ in the southern Galactic
hemisphere, not so far away from the patch of sky we are considering
\citep{djorg,lopez}.
We have checked the warp model adopted by R03 and we found it in excellent
agreement with the observations presented by \citet{lopez}.
\citet{r03} tested their model against observations in several directions and
suggest that it provides reliable predictions down to $V\sim 22$.
In the R03 model the extinction is modeled by a diffuse thin disc. 
In the comparisons with observed
data presented in the following we adjusted the reddening assigned to synthetic
stars by the model to match the observed degree of extinction toward the
considered direction.

\begin{figure}
\includegraphics[width=84mm]{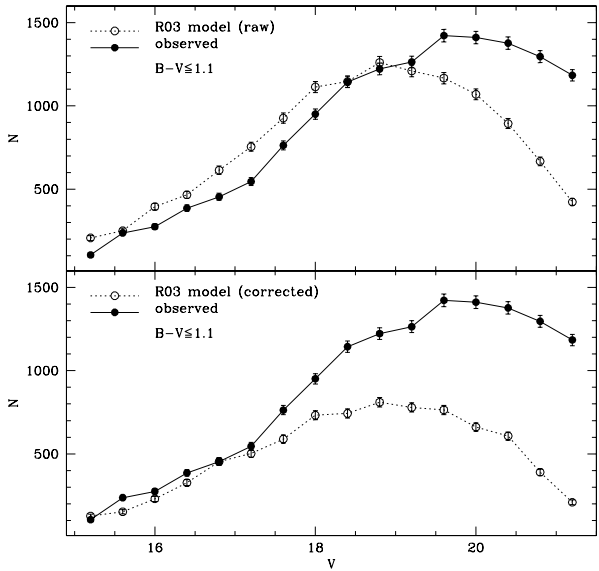} 
\caption{Luminosity Functions of the observed sample (filled circles and
continuous line) and of the synthetic sample (open circles and dotted line).
In the upper panel the raw synthetic sample is considered, in the lower panel 
the synthetic sample has been corrected for incompleteness and normalized as
described in Sect.~2.4 (see also Fig.~3).}
\end{figure}

\begin{figure*}
\includegraphics[width=168mm]{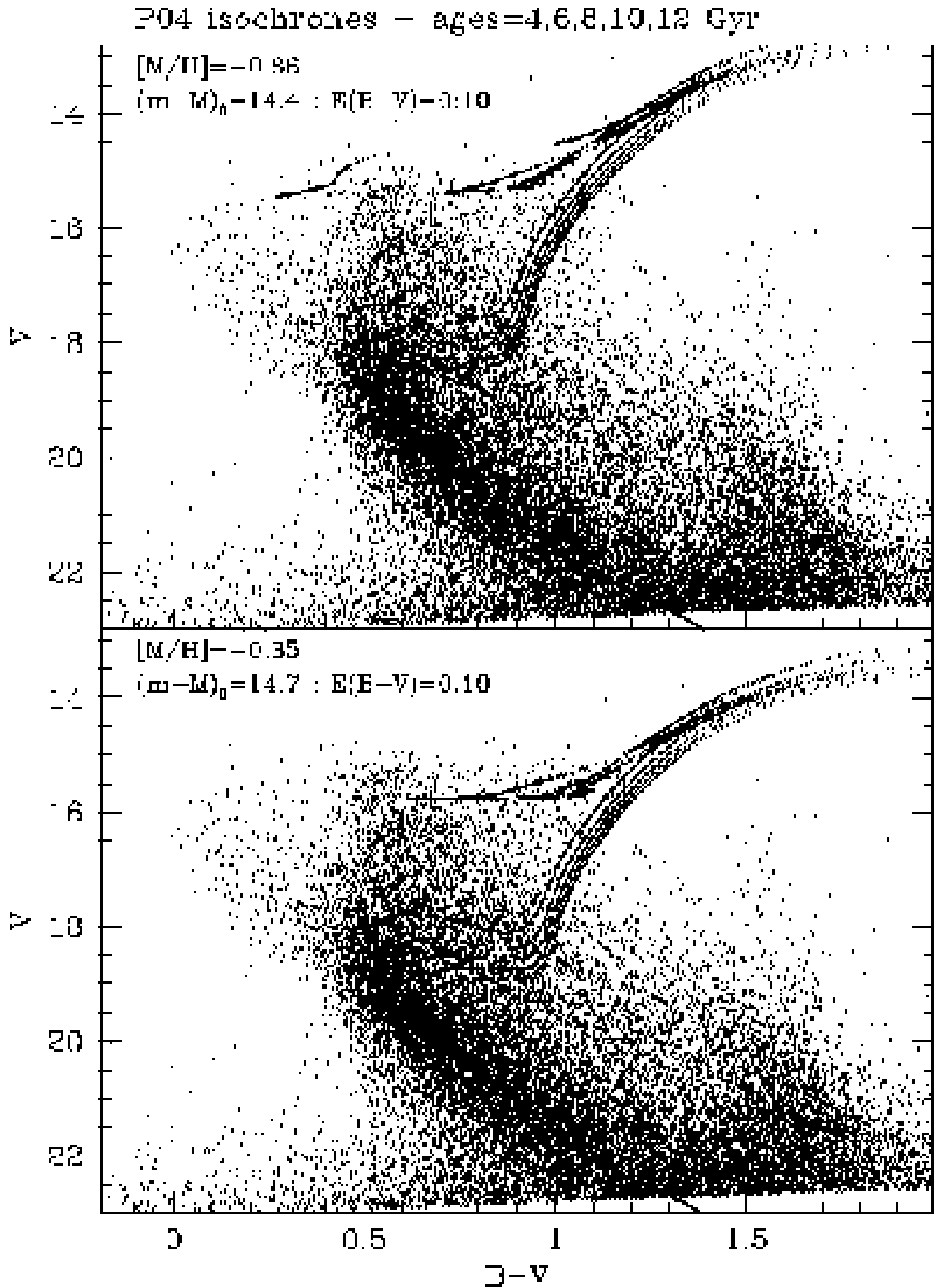} 
\caption{Results of the MS-fitting procedure with isochrones of different
metallicity. Upper panel: $[M/H]=-0.66$; lower panel: $[M/H]=-0.35$.
The isochrones are from \citet{p04}.}
\end{figure*}

In Fig.~3 we compare the observed CMD of F-XMM with the predictions of the R03
models for a field of the same area and in the same direction. The comparison is
limited to the range in which CF$\ge$ 50\% in the observed sample, i.e. to
$V\le 21.5$.
We included the
effect of photometric ``noise'' by adding to each synthetic star a $\delta mag$
extracted at random from a Gaussian distribution having $\sigma$ equal to the
average photometric error of the observed stars at the considered magnitude.

Even from a first glance to the upper panels of Fig.~3 it is quite clear that 
the above-described sequence that dominates the F-XMM CMD has no counterpart in the 
Galactic model. In the latter CMD
the vertical sequence of field MS stars reaches $V\simeq 19.0$ and then
abruptly bends to the red, changing into a sparser sloped band with a sharp 
faint/blue edge.
This is a well known general feature of the CMDs of Galactic fields, 
observed by various authors and predicted by all Galactic models 
\cite[see, for example][]{cast,sdgs1,prandoni,lorenzo,ibaring}.
On the other hand the observed ``unknown'' sequence curves 
gently, always remaining to the blue of the synthetic sequence down to 
$V=21.5$ (and beyond, not shown here), and remains densely clustered all over
its length.
The observed CMD shows also a clear excess of Blue Plume 
(BP; at $0.0\le B-V\le 0.5$ and $16.0\sol V\sol 18.5$) stars with respect
to the synthetic CMD. 

The direct comparison with the raw synthetic CMDs is not quantitatively correct
since it neglects the incompleteness of the observed sample.  
In fact, the synthetic CMD is 100\% complete at all magnitudes. The lower left
panel of Fig.~3 shows the synthetic CMD once the incompleteness is
taken into account. We have associated to each star of the raw synthetic sample 
the joint CF factor of the observed sample at its magnitude. Then we extracted
at random a number in the range [0.0,1.0] and we compared it with the computed
CF. If the random number was smaller than or equal to the characteristic CF of 
the synthetic star the star was retained, otherwise it was excluded from the 
sample.
The overall appearance is now more similar to the observed CMD, 
in terms of number
of stars in regions of the CMD dominated by bona fide field stars. However, the
synthetic CMD still shows an excess of stars of a factor $\sim 1.4$ within the box
superposed on the left-hand panels of Fig.~3, enclosing the brightest part of the
sequence of field MS-TO stars. 
The synthetic M dwarfs brighter than $V\simeq 20.0$ 
are overabundant by the same factor with respect to their observed counterparts.
These comparisons suggest that the model is over-predicting the number of field
stars in this direction. To take into account also this possible effect we
picked up stars at random from the completeness-corrected synthetic CMD until
we obtained a sample having the same number of stars populating the selection box
as the observed one. The CMD of the ``normalized'' sample is shown in the lower
right panel of Fig.~3. This plot provides a realistic view of the contamination
of the observed CMD by field stars. It may be appreciated that the highest degree
of contamination is around the TO of the ``unknown'' sequence, around $B-V\simeq
0.6$ and for $V\le 19.0$. Any eventual Sub Giant Branch of the unexpected
population is completely hidden by field dwarfs having $B-V\ge 0.8$. These facts
constitute the main limitation in obtaining strict constraints on the age of 
the newly-identified population.

In Fig.~4 we compare the LFs of the F-XMM field and of the synthetic sample
for all stars having $B-V\le 1.1$, to exclude the Galactic M dwarfs from the
comparison. In the upper panel of Fig.~4 the observed LF is compared to the raw
model predictions while in the lower panel the comparison is with the LF of the
completeness-corrected and normalized synthetic sample. Fig.~4 shows that there
is a significant excess of stars in the observed CMD corresponding to the unknown
MS population, even if incompleteness and density normalization are not taken
into account. The excess of observed stars arise between $V\sim 18.0$ and $V\sim
19.0$ and persist down to the limiting magnitude of the observed sample (not
shown here). 

From the above comparisons we conclude that in the XMM field there is indeed a 
conspicuous stellar population that is not comprised in the R03 Galactic Model.
We identify this population with the newly discovered Canis Major stellar system,
since the considered field is well within its main body, as described in Pap-I.
Hence we identify the sequence dominating the F-XMM CMD as the MS of the
Canis Major galaxy, revealed for the first time by the observations presented
here.

\subsection{Distance, metallicity and age}

Wetry to constrain the distance to the CMa system in the XMM field by fitting
the observed MS with theoretical isochrones. Here and in the following we will
use isochrones taken from the very recent and complete set of 
\citet[][hereafter P04\footnote{The isochrones can be retrieved 
from http://www.te.astro.it/BASTI/index.php}]{p04}.

The fact that the RGB of the galaxy is dominated by M-giants strongly
suggests a mean metallicity larger than $[M/H]\simeq -1$
\citep{bell03b,rochap,martin}. In particular \citet{crane} derived $[Fe/H]=-0.4$
from low resolution spectra of Ring M-giants in the vicinity of the main body of
CMa. Hence we will adopt isochrones of metallicity $[M/H]=-0.66$ and
$[M/H]=-0.35$ for the present analysis. The results of the MS-fitting procedure
are shown in Fig.~5. The upper panel shows the comparison with isochrones with
$[M/H]=-0.66$, the lower panel shows the comparison with $[M/H]=-0.35$
isochrones.

In both cases the best fit is obtained assuming E(B-V)=$0.10$, in reasonable
agreement with the SFD98 reddening maps (E(B-V)=$0.14$). 
Several authors have suggested that SFD98 reddening maps may overestimate the 
actual reddening by a factor $\simeq 1.14 - 1.35$ 
\citep{stanek,boni,dutra,sumi}. If
this possible correction is taken into account the agreement between our 
best-fit reddening and the estimates of the SFD98 becomes excellent.

The best-fit distance modulus is $(m-M)_0=14.4\pm 0.1$ and 
$(m-M)_0=14.7\pm 0.1$ when the
$[M/H]=-0.66$ and $[M/H]=-0.35$ isochrones are adopted, respectively. 
The value $(m-M)_0=14.4$ corresponds to a heliocentric distance of $D_{\sun}=7.6\pm 0.4$
kpc, in full agreement with the estimate of Pap-I, within the uncertainties
($D_{\sun}\simeq 7.1$ kpc). Similarly, $(m-M)_0=14.7\pm 0.1$ corresponds to
$D_{\sun}=8.7\pm 0.4$ kpc, $\sim 20$\% larger than the estimate of Pap-I but
still fully compatible with it, given the $\sim 30$\% systematic uncertainty
affecting the M-giants distance scale adopted in that work \cite[see][for
discussion]{martin,newberg2}.

No stringent constraint on metallicity can be derived from the comparisons 
shown in Fig.~5. We just note that the lower distance modulus obtained with the
$[M/H]=-0.66$ isochrones put the theoretical RC locus just above the saturation
limit of our photometry, in better agreement with observations compared to
the $[M/H]=-0.35$ MS-fitting solution. It has also to be recalled that
theoretical isochrones predict the location of the Zero Age Horizontal Branch,
i.e. the faint edge of real distributions of Helium Burning stars. 

A significant lower limit to the age of the main population 
can be obtained by comparison with isochrones with the bright end of the
observed MS. From Fig.~5 we obtain an
age $\ga 6$ Gyr for $[M/H]=-0.66$ or an age $\ga 4$ Gyr for $[M/H]=-0.35$.
On the other hand the corresponding upper limit is made greatly uncertain by
the high degree of field contamination in this portion of the CMD. Ages as large
as 10 Gyr cannot be excluded with the present data. Independent of the quoted
uncertainties it should be also considered that (a) the CMa system probably hosts
stars of various ages and metallicities, as most of the local dwarf galaxies 
actually do \citep{mateo}, and (b) the galaxy has a sizable dimension along
the line of sight ($\sigma_{(m-M)_0}\sim 0.5$ mag, according to Pap-I), hence the
observed CMD sequences may appear wider and less well-defined than in
smaller or more distant systems just because
of this differential-distance effect. Finally, to fully populate the observed
Blue Plume, stars as young as $\sim 1$ Gyr should also be present in CMa,
suggesting a recent (minor) burst of star formation, similar to what observed 
in the Sgr dSph \citep{sdgs2,ls00}.

From the above comparisons it appears clear that more complete photometry
(reaching the Tip of the RGB) and a suitable (observed) control field to perform
a robust field decontamination are mandatory to obtain a deeper insight into the
physical characteristics and the stellar content of the CMa system. Nevertheless,
we have provided here the first clear detection of the MS of the galaxy, an
independent estimate of its distance, in good agreement with the results of
Pap-I, as well as the first broad constraint on the age of the main stellar
populations of the system. In the following we will use the
results obtained in this section as guidelines for the analysis of the CMD of
populations possibly associated with the CMa/Ring system in the background of
Galactic open clusters.

\section{CMa in the background of Open Clusters?}

We searched the WEBDA database for published CCD photometry of open clusters
lying in the broad region covered by the CMa galaxy, i.e. $220^o\le l \le
260^o$ and $-20^o\le b\le 0^o$, according to Pap-I. 
We found a handful of possibly useful
datasets (with sufficiently deep CCD data over a field of at least a few 
arcmin across), that we briefly analyze in the following. 
The considered photometries, being optimized to image nearby clusters, are too 
shallow to provide firm detections (and/or {\em non-detections}) of the CMa
population. Uncertainties in the reddening may also affect the analysis of these
fields.
Nevertheless some interesting clues of the presence of CMa stars can
still be found and may provide useful hints for future observations.

\subsection{The background of NGC~2477}

The open cluster NGC~2477 is located at (l,b)=(l=$253.6^o$; b=$-5.8^o$).
Its age is $\sim 1$ Gyr, its heliocentric distance is $\simeq 1.3$ kpc 
\citep{kas}, and the metallicity is $[Fe/H]\simeq -0.5$
\citep{friel}. Unfortunately, the field of NGC~2477 is
strongly affected by interstellar extinction ($\langle E(B-V)\rangle=0.77$ 
according SFD98),
whose actual amount is very uncertain \cite[see][]{hart,kas} and varies
significantly across the observed field (with a standard deviation of 0.07 mag,
SFD98). This constitutes the major source of uncertainty for the results
obtained in the following.

\citet{moma} recently presented the B,V photometry of a wide
field ($\simeq 34^\prime \times 33^\prime$) centered on the cluster.
The CMD obtained by \citet{moma} is shown in the upper left panel of Fig.~6.
The cluster MS is a narrow diagonal
band crossing the CMD from $B-V\simeq 0.5$, $V\simeq 13.0$ to
$B-V\simeq 2.0$, $V\simeq 22.0$. The cluster RC is clearly
visible at $B-V\sim 1.3$, $V\simeq 12.5$.
The most striking feature of the
diagram is the large number of stars that are located along a wide sequence
paralleling the cluster MS from $B-V\sim 1.0$, $V\sim 18.0$ down to
$B-V\sim 1.5$, $V\sim 22.0$.
Indeed the feature is reminiscent of the upper
MS - Turn Off (TO) region of an intermediate-old stellar population, like that
described in the previous section and attributed to the CMa system. Moreover,
from the bright head of this putative sequence departs a Blue Plume of stars
($0.5\le (B-V)\le 1.0$, $17.0\sol V\sol 20.0$) similar to what is observed
in the XMM field described in the previous section.

\begin{figure}
\includegraphics[width=84mm]{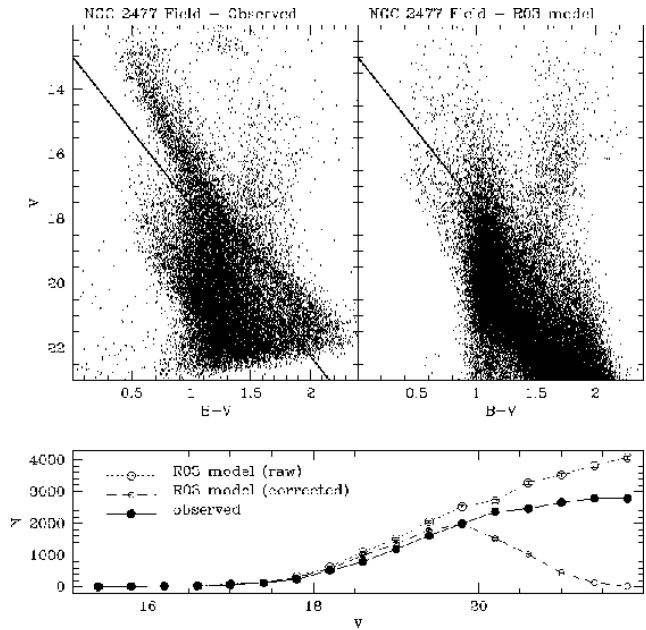} 
\caption{Upper panels: observed CMD of the NGC~2477 field (left panel)
and the corresponding synthetic CMD from the R03 model (right panel).
The selection line separates the background MS sequence form the bulk of the
cluster stars. Lower panel: comparison between
Luminosity Functions. The symbols are the same as in Fig.~4 except for the
smaller open circles that represent the synthetic LF once corrected for the
completeness function of F-XMM shifted by $-1.0$ mag to account for the
different limiting magnitudes of the two samples.}
\end{figure}

The upper right panel of Fig.~6 shows the corresponding synthetic CMD from the R03 model. The
photometric uncertainties in the synthetic CMD have been implemented as described
in Sect.~2.4, above. The diagonal line over-plotted on both the observed and
synthetic CMDs is intended to separate the background MS population we are
interested in from the bulk of the cluster stars.
In the present case the synthetic CMD is quite similar to the
observed one, if we neglect the sequences associated with the cluster,
suggesting that the CMD is dominated by Galactic field population. However we
note that (a) the actual morphology of the main observed sequence is different
from what predicted by the model, (b) the BP population is much richer in the
observed CMD, reaching also fainter magnitudes, (c) a sparse clump around
$V\simeq 17.5$ and $B-V\simeq 1.6$ in the observed CMD seems to have no
counterpart in the synthetic one (see below). 

In the lower panel of Fig.~6 the LF of the stars lying to the blue of 
the selection line are compared. A slight excess in the synthetic LF is apparent
for $V>18.0$, growing toward fainter magnitudes and reaching a factor $\sim 1.4$
at the limiting magnitude of the observed CMD. However, it has to be
considered that while the synthetic sample
is 100\% complete at any magnitude the observed sample should be significantly
incomplete for $V>20$, i.e. within the last two magnitudes from the faint limit 
of the photometry. Under similar conditions (i.e. same instrument, similar
seeing, similar exposure times) the completeness of the F-XMM sample is always
lower than 50\% in the same range near the faint limit of the photometry.
If we correct the synthetic LF for completeness (as done in Sect.~2.4, above)
adopting the Completeness Function of F-XMM (shifted by $-1.0$ mag to partially
account for the different limiting magnitudes, i.e. $V\simeq 22.0$ for NGC~2477
and $V\simeq 23.5$ for F-XMM), the synthetic LF (small open circles)
falls well below the observed one for $V>19.5$. Hence, the test concerning the MS
populations is not conclusive.

The case for the evolved population and for the Blue Plume is more interesting
and it is briefly discussed below in comparison with the arguments provided by
M04 to explain the population in the background of NGC~2477 as completely due to
the Galactic warp. M04 made the same comparison shown in Fig.~6 here but they
selected from the synthetic catalogue only the stars with heliocentric distance
between 7 and 9 kpc, claiming that all the sequences possibly attributable to
CMa (including the BP and the RC features) ``are fully reproduced by the
Galactic warped model'' (R03). The comparisons shown in Fig.~7 below demonstrates
that this is not actually the case.

\begin{figure}
\includegraphics[width=84mm]{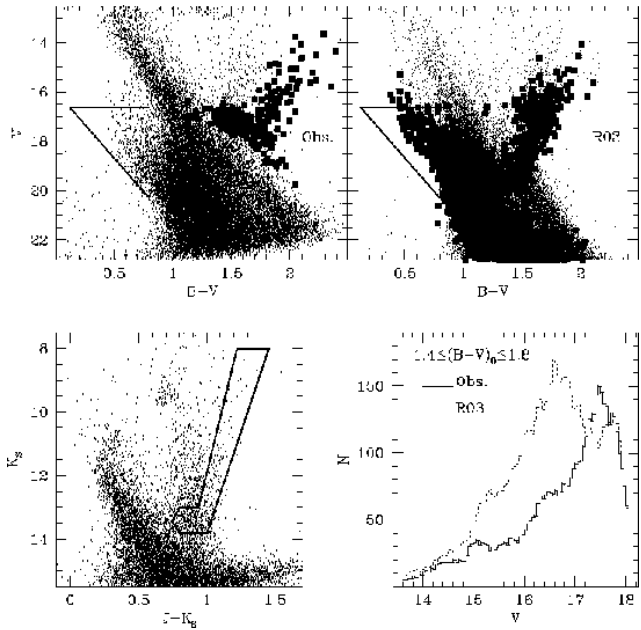} 
\caption{Upper left panel: observed CMD of NGC~2477. The heavy dots are the RGB
and RC stars selected in the infrared CMD shown in the lower left panel.
Upper right panel: synthetic CMD of an equivalent area corrected for extinction
with the same assumptions adopted by M04. The heavy dots are the synthetic stars
with heliocentric distance between 7 and 9 kpc (as in M04). Lower right panel:
smoothed histograms of the stars in the color range of the Red Clump.
}
\end{figure}

The upper panels of Fig.~7 shows once again the comparison between the observed
CMD of NGC~2477 and its synthetic counterpart from R03 \footnote{Actually a
different realization with respect to that shown in Fig.~6. To check the effect
of statistic noise on simulated CMD we analyzed four different realizations of
the same synthetic sample in the present case. The differences among the various
realizations are negligible under any relevant aspect.}.
We corrected the synthetic CMD with the same assumptions on reddening adopted by
M04 \cite[$<E(B-V)>=0.52 \pm 0.04$, after][]{boni}. 
In the observed CMD we
plot as heavy dots the putative RC and RGB stars of CMa we have selected from
the Near Infrared CMD of the same region that is shown in the lower left panel
of Fig.~7. The RGB sequence is readily identified in this diagram and also the
RC peak is more evident (see also Pap-I and references therein). The triangular
box encloses the bluest part of the BP distribution (which extends to $B-V\sim
1.0$ in the observed CMD). In the synthetic CMD we plotted as heavy dots the
stars fulfilling the same selection criteria adopted by M04 ($7<d_{\odot}<9$
kpc).

The overall morphology of the observed CMD is reasonably reproduced
by the distance-selected synthetic sample, though there are 
still significant differences:

\begin{enumerate}

\item{} The BP sequence is not correctly reproduced in position and in number.
The observed stars falling in the selection box are 582 while the synthetic
stars are just 285, a full 10 $\sigma$ difference. Moreover the faint envelope
of the synthetic BP sequence is one full magnitude brighter than the
corresponding feature in the observed diagram.

\item{} The selection of synthetic stars in heliocentric distance indeed
show evidence for a Red Clump population. However, as clearly shown by the LFs in
the lower right panel of Fig.~7, the synthetic RC peak is more than $0.8$ mag
brighter than the observed one. Note that the positions of the peaks are
unaffected by the inclusion of photometric errors in the synthetic sample.
The synthetic RC appears more clumpy and well-defined in the plots produced by
M04 (see their Fig.~3) because they neglect the effects of observational errors,
but the average magnitude of the peak is the same as found here.
Moreover, the overall comparison between the RC LFs suggests that the model
is seriously overestimating the total number of stars in the field (as for
F-XMM, see Sect.~2.4), providing a further reason to reconsider the comparison
of MS LFs shown in Fig.~6.

\item{} M04 found that the average metallicity of the selected synthetic sample
(heavy dots) is $[Fe/H]=-0.45\pm 0.25$ and the average age is $5.0\pm 1.5$ Gyr,
similar to what is found for the CMa system. We verified these results with four
different realizations of the synthetic sample of the NGC~2477 field and we 
fully confirm the metallicity result. However, it transpired that M04
misinterpreted the column of the output files of the R03 model labeled ``age''
as providing the age in Gyr of any given star. In fact this is an integer code
number that classifies the stars according to the Galactic component they belong
to (thin disc, thick disc, bulge and halo) and to age ranges, within the thin
disc population (age/population class). Due to an incredibly lucky coincidence the
average of the  age/population class and of the actual age of the selected
sample are the same (i.e. $5$, a pure number, and $5$ Gyr, respectively) but 
the standard deviations are significantly different ($1.5$ and $3$ Gyr, 
respectively). A closer look at the age/population class distribution shows that
$\simeq 45$\% of the selected stars have age $< 4$ Gyr, hence not similar to
the typical CMa population, and $\simeq 10$\% are thick disc stars, hence not
possibly related to the warp. In conclusion, only less than 50\% of the
selected sample claimed to reproduce the CMa features can be realistically
related to it.

\item{} Finally, we note that there is no reasonable reddening value
that may simultaneously allow a reasonable match to {\em all} the main features
of the synthetic and observed CMDs of Fig.~7 (i.e. MS, BP and RC).

\end{enumerate} 

In conclusion the Galactic warped model adopted by M04 completely fails to
reproduce the observations for what concerns the Blue Plume and the Red Clump
in the background of NGC~2477. The possibility that these stars are associated
with the CMa system or its debris (Ring) remains fully open. Deeper observations
and extensive artificial stars tests are needed to verify the existence of 
their MS counterparts at $V>18.0$.

\subsection{The background of Tombaugh~1 and Berkeley~33} 

The open cluster Tombaugh 1 (Tom 1) is located at (l,b)=($232.2^o,-7.3^o$).
According to \citep{tom1} the most likely metallicity is $[M/H]\sim0.0$, the age
is $\sim 1$ Gyr and it is located at $\sim 3$ kpc from the Sun [$(m-M)_0\sim
12.4$]. In the upper panels of Fig.~8 the CMD of Tom~1, as derived by 
\citep{tom1}, is compared to the corresponding synthetic CMD from the R03 
model. The observed CMD is dominated by the cluster MS going from 
$V-I\simeq 0.7$ and $V\simeq 14$ to $V-I\simeq 2.0$ and $V\simeq 21.5$. 
Note that a broad and sparse parallel sequence seems to appear to the blue of 
the cluster MS (and of the selection line introduced in the previous section)
at $V\ge 18.5$. The sequence seems to have no clear counterpart in the synthetic
CMD and resembles the upper-MS of CMa as identified in F-XMM.  
The LFs shown in the lower panel of Fig.~8 confirms
this view. The selected stars (i.e. the stars bluer than the selection line)
are more numerous in the observed sample compared to the synthetic one all
over the considered range of magnitudes ($19.5\le V\le 21.5$). Also in this case
the observed sample should be drastically pruned by incompleteness in the
considered range, so the real overdensity should be significantly larger than
what is reported in Fig.~8.

\begin{figure}
\includegraphics[width=84mm]{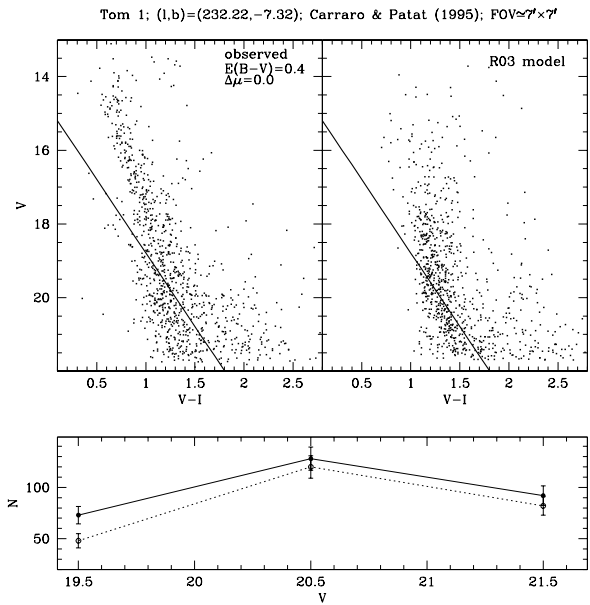} 
\caption{Comparison of the observed and synthetic CMD (upper panels) and LFs
(lower panel) for the field of Tombaugh~1. The symbols are the same as in
Fig.~6.}
\end{figure}

The same comparison is shown in Fig.~9 for the field of
the open cluster Berkeley~33 (Be~33).
Be~33  is located at (l,b)=($225.4^o,-4.6^o$). It has an age of
$\simeq 0.7$ Gyr and $[M/H]\sim -0.7$; its distance from us is $\simeq 3.7$ kpc
\cite[i.e., $(m-M)_0\simeq 13.3$,][]{be33}.
The cluster MS crosses the diagrams from 
$B-V\simeq 0.6$ and $V\simeq 13$ to $B-V\simeq 1.4$ and $V\simeq 21$.

Also in this case a wedge-shaped sequence possibly corresponding to the upper-MS
of CMa is observed to the blue of the selection line and for $V\ga 19.0$.
In this case the LFs of the observed and synthetic sample,
shown in the lower panel of Fig.~9, are essentially indistinguishable. Again,
the high degree of incompleteness that is expected to affect the observed sample
for $V\ge 20.0$ suggests the presence of an excess of stars not predicted by the
Galactic model. 

More complete and deeper photometry of wider fields
surrounding the clusters are needed to confirm the tentative detections
described above. 

\begin{figure}
\includegraphics[width=84mm]{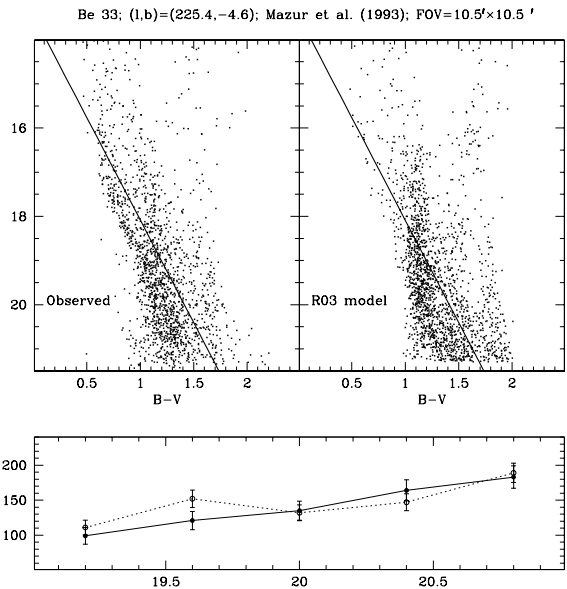} 
\caption{Comparison of the observed and synthetic CMD (upper panels) and LFs
(lower panel) for the field of Berkeley 33. The photometry is from 
\citet{be33}. The symbols are the same as in Fig.~6 and Fig.~8.}
\end{figure}

\subsection{Fields at $b<-14\degr$}

Finally, Fig.~10 shows the CMDs of three OCs having $220\degr\le l\le 260\degr$
and $b<-14\degr$, namely NGC~2204, Melotte~66 \citep{kas}, and
NGC~2243 \citep{2243kal}. 
In spite of the larger field of view with respect to the
observations of Tom~1 and Be~33 described above, there is no sign of a defined
upper-MS sequence to the blue of the selection lines in these CMDs. The
assessment of the significance of a {\em non-detection} is a prohibitive task
given the available observational material. The comparison with the CMDs of 
Tom~1 and Be~33 (note that NGC~2204 is essentially at the same galactic 
longitude as  Be~33) simply suggests that the density of CMa stars is 
significantly decreased going from $-7\degr \le b \le -4\degr $ to $b<-14\degr$.

\begin{figure}
\includegraphics[width=84mm]{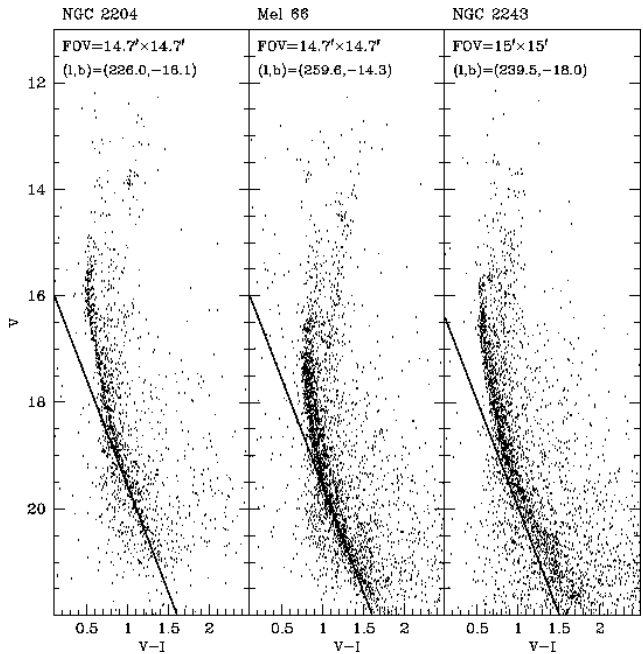} 
\caption{CMDs of open clusters having $220\degr\le l\le 260\degr$ and
$b<-14\degr$. The data for NGC~2204 and Melotte~66 are from \citet{kas},
the data for NGC~2243 are from \citet{2243kal}.}
\end{figure}

From the analysis presented in this section we simply conclude that the
existing CMDs sampling the stellar population in the background of clusters
located in the same broad region of the sky that hosts the CMa system are
compatible with the results presented in Pap-I regarding the overall structure of
the galaxy. As a guideline for future observations it is also suggested that
the additional effort required to observe wider fields and to obtain deeper
photometry of clusters in this region may be worth doing and highly rewarding 
in terms of scientific returns, and it may also help to obtain a more thorough
understanding of the population of the clusters themselves.

\begin{table*}
 \centering
 \begin{minipage}{140mm}
  \caption{Selected Fields.}
  \begin{tabular}{@{}lccccc@{}}
  \hline
Field Name & l & b & r & $<E(B-V)>$\footnote{Average reddening $\pm$ standard
deviation from star-by-star interpolation on the SFD98 maps.} & Data Source \\
  \hline
CMa   &$238\degr \le l \le 244\degr$ &$-11\degr \le b \le -6\degr$&~--~ &$0.27\pm
0.09$& 2MASS - PSC\footnote{Point Source Catalogue}\\
CMa-CF &$238\degr \le l \le 244\degr$ &$+6\degr \le b \le +11\degr$&~--~ &$0.12\pm
0.02$& 2MASS - PSC\\
WARP   &$272\degr \le l \le 278\degr$ &$-11\degr \le b \le -6\degr$&~--~ &$0.29\pm
0.05$& 2MASS - PSC\\
WARP-CF   &$272\degr \le l \le 278\degr$ &$+6\degr \le b \le +11\degr$&~--~
&$0.23\pm 0.04$& 2MASS - PSC\\
&&&&&\\
CMa (GSC2) &$l=243\degr$ & $b=-8\degr$ & $r=1.0\degr$&$0.18\pm 0.02$ &
GSC2.2\footnote{Only entries classified as stars have been retained in all the
samples extracted from the GSC2.2 catalogue.}\\
CMa-CF (GSC2) &$l=243\degr$ & $b=+8\degr$ & $r=1.0\degr$&$0.12\pm 0.01$ &GSC2.2\\
WARP (GSC2) &$l=274\degr$ & $b=-8\degr$ & $r=1.0\degr$&$0.27\pm 0.04$ &GSC2.2\\
WARP-CF (GSC2) &$l=274\degr$ & $b=+8\degr$ & $r=1.0\degr$&$0.21\pm 0.05$ &GSC2.2\\
\hline
\end{tabular}
\end{minipage}
\end{table*}


\section{The CMa overdensity and the Galactic Warp}

Reconsidering the results of Martin et al. (2004a), M04 studied the North/South asymmetry 
in starcounts of
M-giants in the latitude stripe $|b|<20\degr$ and in the latitude range 
$235\degr \le l\le 245\degr$ and showed that in this spot of the sky the CMa
overdensity is completely erased if (i) a model warp of amplitude 
$w=2.0\degr \pm 0.3\degr$ is adopted {\em and} (ii) the correction introduced 
by \citet{boni} is applied to the SFD98 reddening values. It is quite obvious 
that any overdensity in the proximity of the Galactic Plane can be locally
reproduced  by a particular model of the warp. 
What should be demonstrated is that a global
model of the warp, broadly reproducing the observations in the whole Galaxy, is
also able to reproduce the CMa overdensity.\footnote{It should be noted that the
warp model used by M04 is so crude that can be used only on small scale, by
definition. To emulate the local effects of the Galactic warp they adopt 
$b=-2\degr$ as the plane of symmetry of the Galaxy, instead of $b=0\degr$. In
\citet{martinb} we show that the CMa overdensity is detected even using the M04
warp model, once the distance information is properly taken into account.}
In fact this is not the case with
the only global model adopted by M04, i.e. the warped and flared Galactic 
model by R03 they use for comparison with the NGC~2477 field. Even accepting
that this model reproduces the observations in the latter field (that is far
from proven, see Sect.~3.1 above) {\em it completely fails to reproduce the
morphology of the CMD and the star counts in the XMM field}, as exhaustively
shown in Sect.~2.4, above. There, a very significant unexplained overdensity and
stellar population remains, in excess of the warp model encapsulated in the
R03 model. Moreover in their analysis M04 completely neglect the distance
information provided by the photometric parallax of M-giants that is one of the
key elements of the discovery presented in Pap-I. In the following we shall
demonstrate that the consideration of this additional information easily singles
out the CMa overdensity from the background of the Galactic warp.

While extensively checked with various sets of observations the R03 model cannot
provide a completely safe comparison and the ideal testbed to proof the reality
of the CMa structure is the {\em real} Galaxy. In the following subsections we
perform direct comparisons with suitable control fields using data from 2MASS 
and from the 
GSC2.2\footnote{See http://www-gsss.stsci.edu/gsc/gsc2/GSC2home.htm} surveys.

\subsection{CMa is not the warp: 2MASS data}

All the tracers adopted in the literature to study the Galactic warp agree in
placing the maximum amplitude of the southern Galactic warp around $l=270\degr$,
the North/South asymmetry remaining near the maximum in the broad range
$230\degr \le l\le 280\degr$ \citep{yusi,lopez}. 
The classical parameter adopted to
quantify the asymmetry produced by the warp at a given ($l;|b|$) is the ratio
between the counts (of any given density tracer) in the northern Galactic 
hemisphere and the counts in the corresponding region in the
southern hemisphere ($l;-|b|$), $R=N_{North}/N_{South}$. 
At the longitudes of the
maximum amplitude of the warp, R is $\sim 0.7$ in the southern warp 
\citep{yusi,lopez}. Here, since we deal with southern structures we
will use $R\prime={1\over{R}}=N_{South}/N_{North}$. The typical value of 
$R\prime$ around the southern maximum is $R\prime\sim 1.4$, which we will adopt as
a reference value in the following. In any case values of $R\prime$ significantly
larger than 1.0 indicate an asymmetry between the Galactic hemispheres, in
particular an excess of star counts in the Southern hemisphere. Note
that computing $R\prime$ from a homogeneous dataset (as, for instance 2MASS)
removes 
most of the problems associated with the incompleteness of the samples since the
parameter is the {\em ratio} of starcounts from samples with very similar
completeness as a function of magnitude.
 
\begin{figure}
\includegraphics[width=84mm]{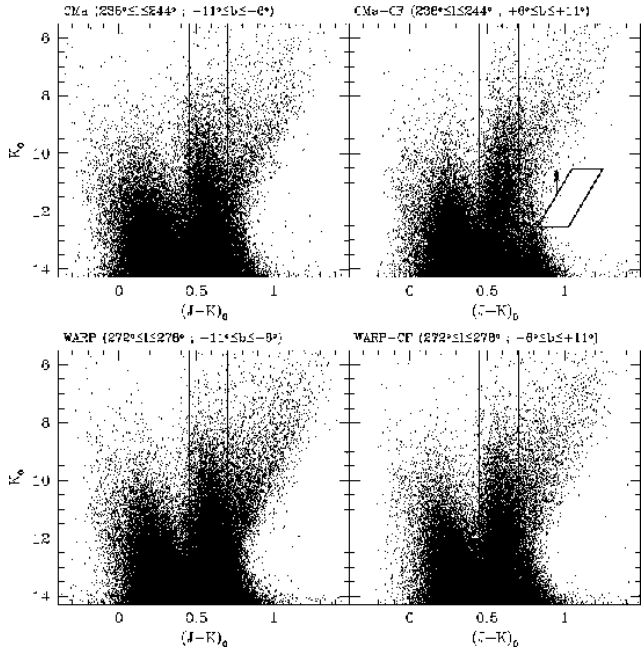} 
\caption{NIR CMDs of the 2MASS selected fields described in Tab.~2. The vertical
lines encloses the color range populated by Red Clump stars at any distance. The
adopted M giants selection box is plotted in the upper right panel. The arrow
indicates that the box is shifted along the magnitude axis to count M giants as
a function of $K_0$.}
\end{figure}

The key point of the following analysis is that if the CMa overdensity is due to
the southern Galactic warp we shall observe similar overdensities and
populations also around $l=270\degr$ where the warp asymmetry should be similar
if not larger. 
To do this comparison we carefully selected two windows of low
reddening $E(B-V)<0.5$ and of low average reddening 
$\langle E(B-V)\rangle <0.3$ and their
respective control fields in the northern hemisphere with the same reddening
constraints. The data have been taken from 2MASS adopting the same selections
as in \citet{bell03b}.
The chosen windows - of the same area and covering the same
latitude ranges - are described in Table 2. For brevity the fields centered at
$l=241\degr$ have been dubbed CMa (South) and CMa-CF (North), those centered at
$l=275\degr$ are called WARP (South) and WARP-CF (North). The data have been
corrected for extinction according to the SFD98 maps. We note that all the
results presented in this section remain valid also if the recipe by 
\citet{boni} is adopted. The CMDs of the four considered fields are shown in
Fig.~11. The vertical lines enclose the color range populated by RC stars at
various distances from us. The parallelogram plotted in the upper right panel of
Fig.~11 is the same selection box adopted by \citet{bell03b} to search for
overdensities of M-giants as a function of distance. The box, shaped around
the RGB of the Sgr dSph galaxy, is shifted along the magnitude axis counting the
stars that fall into it as a function of the shift. The shifts are converted to
distance moduli assuming $(m-M)_0=16.90$ for Sgr, as done by
\citet{majewski} and \citet{martin}.

\begin{figure}
\includegraphics[width=84mm]{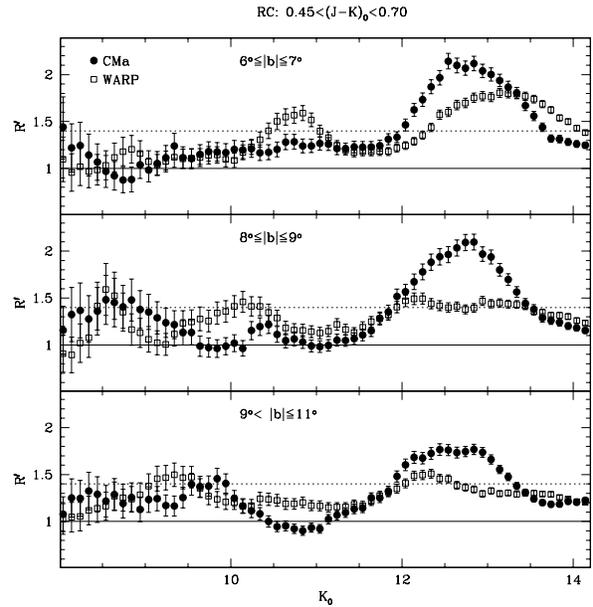} 
\caption{R$\prime$ distributions of RC stars in the CMa (filled circles) and
WARP (open circles) fields in different ranges of Galactic latitudes.
The dotted line marks the level of asymmetry typical of the Galactic warp,
$R\prime 1.4$ \citep{yusi,lopez}.}
\end{figure}

In Fig.~12 the distributions of $R\prime$ as a function of $K_0$ magnitude and
computed for stars in the color range of the Red Clump are plotted for different
ranges of Galactic latitudes. In the range $6\degr \le |b|\le 7\degr$ a strong
peak in $R\prime$, reaching $R\prime=2.1$, is clearly detected in the CMa
direction at $K_0\sim 12.7$. According to the model by \citet{sg02}, the
absolute K magnitude of the Red Clump of a population having 
$[M/H]\simeq -0.4/-0.7$
in the age range $4-10$ Gyr is $<M_K>\simeq-1.5\pm 0.2$. 
Hence the peak observed in
the CMa direction corresponds to an overdensity at the distance $(m-M)_0\simeq
14.2 \pm 0.4$, in good agreement with the distance estimates for the CMa galaxy
obtained in Pap-I and in Sect.~2.5, above. 
Note also that the Full Width at Half
Maximum of the $R\prime$ peak is fully consistent with the FWHM of the CMa
structure obtained in Pap-I \cite[see Fig.~6, in][]{martin}.
On the other hand in the WARP direction,
while $R\prime >1$ essentially everywhere, as expected, a much weaker peak is
detected around $K_0\sim 13.1$. Let us assume that both the above described
peaks are due to the Galactic warp. If we imagine the Galactic disc as a series
of concentric circular rings with the outermost rings tilted  
\cite[because of the warp, see][and references therein]{binney} the structures 
at $l\sim 241\degr$ (CMa) and at $l\sim 275\degr$  (WARP)
should lie at the same Galactocentric distance, i.e. in the same {\em ring}.
Seen from the Sun, this ring should appear more distant in the WARP direction,
as observed, given the fainter RC peak detected in the WARP field.
However, to obtain the observed difference in the magnitude of the peaks
($\Delta K_0\simeq 0.4$ mag) the warped ring should lie at a Galactocentric
distance of $\sim 20$ kpc, implying a distance of the CMa structure from the Sun
of $\simeq 15$ kpc, in sharp contrast with the observations (see Pap-I and
Sect.~2.5, above). On the other hand, assuming a realistic Sun-CMa distance 
of $8.0$ kpc, the difference between the two peaks should be $\simeq 0.9$ mag,
much larger than observed. The only way to save the original hypothesis, that
the overdensities in the CMa and WARP directions belong to the same warped ring
of the Galactic disc, is to admit that such a ring {\em is not circular}, but
significantly elliptical instead (axis ratio $\simeq 1.4$). 
Even in this (very unlikely) case it
would be very hard to explain the large difference in the amplitude of the
$R\prime$ peaks, much higher toward $l\sim 241\degr$ than toward 
$l\sim 275\degr$.

\begin{figure}
\includegraphics[width=84mm]{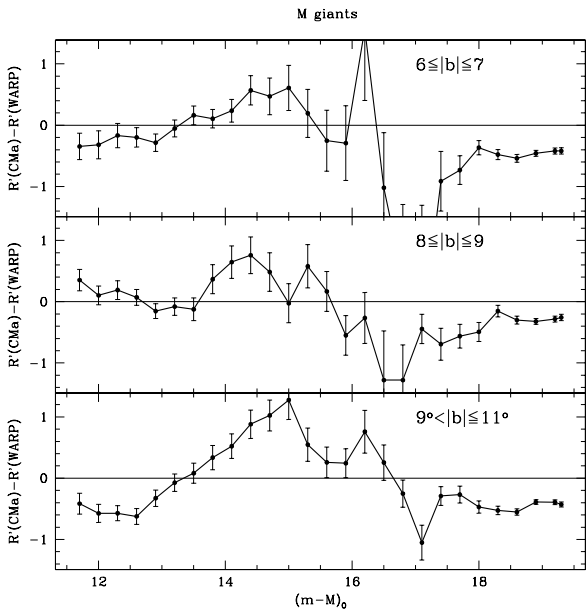} 
\caption{R$\prime$ differences from M giants as a function of distance modulus
 (see Bellazzini et al.(2003b) for details on the adopted technique).}
\end{figure}

Even more interesting, in this sense, is the behaviour of the $R\prime$
distributions at $8\degr \le |b|\le 9\degr$ (Fig~12, middle panel). Here there is no
more peak in the WARP direction. For $K_0>12.0$ the $R\prime$ distribution of
the $l\sim 275\degr$ field settles at the typical ``warp'' value (dotted line)
while the CMa peak is as strong and well defined as at lower latitudes and in
the same position. This comparison alone shows that (a) toward 
$(l;b)\simeq (240\degr;-8.5\degr)$ the South/North asymmetry in RC stars is
significantly larger than in the WARP field (a difference larger than
$7 \sigma$ at $K_0\simeq 12.7$) and (b) the excess of counts in the CMa
direction is confined to a well defined range of magnitudes corresponding to the
distance of the CMa galaxy as obtained in Pap-I and in Sect.~2.5, here.
It should be concluded that while around $l\simeq 275\degr$ the distribution of
RC stars is fully consistent with what is expected from the Galactic warp
\cite[see][]{lopez}, in the $l\simeq 240\degr$ direction a clear and unexpected
overdensity is observed at the distance of $\sim 8$ kpc, fully consistent with
the results of Pap-I. It should be noted that the overdensity detected in the 
CMa direction is not only a South/North excess of starcounts in a limited range
of distances but it is a clear excess {\em with respect to what observed in the
real warp}, i.e. in the $l\simeq 275\degr$ direction. The result is confirmed
also in the latitude range $9\degr \le |b|\le 11\degr$, where the CMa peak
begins to weaken, as expected (see Fig.~6 of Pap-I) while the WARP distribution
is essentially unchanged 
\cite[in agreement with the results by][see their Fig.~15]{lopez}.

Finally, in the hypothesis that the South/North asymmetries observed in the CMa
and WARP fields are both due to the Galactic warp, one can object that since the
warp ring has a larger heliocentric distance in the $l\simeq 275\degr$ direction
than toward CMa, the same $|b|$ range samples different ranges in Z (linear
distance from the Galactic plane) in the two directions. It can be conceived
that a comparison of the $R\prime$ distributions at the same Z is more appropriate
than the comparisons shown in Fig.~12. Let us consider two possible scenarios:
\begin{enumerate}

\item{} If we assume that the warp lies on a
circular ring at the same Galactocentric distance as the {\em observed} CMa
overdensity then the $9\degr \le |b|\le 11\degr$ range toward CMa samples the
same range in Z than the $6\degr \le |b|\le 7\degr$ range in the WARP direction.
Comparing the CMa distribution in the lower panel of Fig.~12 with that of the
WARP field in the upper panel of the same figure it can be noted that the
amplitude of the peaks is similar (the CMa peak being still the strongest of the
two) but the difference in magnitude between the peaks is again $0.4$ mag,
which would force the conclusion that the warp is elliptical 
(already discussed above).

\item{} If we assume that the real difference in distance modulus 
between the warp as seen toward $l\sim 241\degr$ and toward $l\sim 275\degr$
is $0.4$ mag, as indicated by the position of the $R\prime$ peaks in the upper
panel of Fig.~12, then the comparison at the ``same Z'' should be done between
the CMa distribution at $8\degr \le |b|\le 9\degr$ and the WARP distribution at 
$6\degr \le |b|\le 7\degr$. Since the CMa $R\prime$ distribution
is essentially unchanged in the two
latitude strips (upper and middle panels of Fig.~12) this hypothesis must be
rejected: the CMa peak is much higher and at a different Galactocentric distance
with respect to the peak at $l\sim 275\degr$ (possibly) attributable to the 
galactic warp. Note that this peak is seen only at $6\degr \le |b|\le 7\degr$,
while it is completely lacking at higher latitudes.

\end{enumerate}

Hence, the consideration of this possible effect is of no help to the M04
hypothesis: the observed $R\prime$ distribution of RC stars in the CMa field
cannot be reconciled with  the $R\prime$ distribution in the {\em real} warp 
(WARP field).

Essentially the same results are obtained from M giants, as shown in Fig.~13.
In this case, since the tracers are much less numerous than RC stars and the
signal-to-noise ratio of any structure is consequently lower, we had to adopt a
coarser grid to compute $R\prime$ and the final result is best appreciated when
shown as the difference between the $R\prime$ distributions in the considered
directions ($R\prime(CMa) - R\prime(WARP)$) vs. $(m-M)_0$). 
It is clearly evident
that the only significant excess is the broad peak in the range 
$14.0\le (m-M)_0 \le 15.0$ that is detected in all of the considered latitude
ranges. Note that Fig.~13 is directly showing excesses in star counts in the CMa
direction {\em with respect to the WARP direction}, hence the CMa clump of stars
clearly emerges {\em over} the South/North overdensity due to the warp.

It must be concluded that the observed $R\prime$ distributions of RC stars
and M giants
toward $l\sim 241\degr$ and toward $l\sim 275\degr$ are in clear contrast with
the claim that the overdensity discovered in Pap-I and studied here is due
to the Galactic warp (M04). On the other hand, the observational evidence
confirms that around $(l;b)\sim(240\degr;-7.5\degr)$, at $\sim 8.0$ kpc from the
Sun, there is a previously unknown stellar system that is clearly detected even
above the underlying asymmetry due to the Galactic warp: the Canis Major galaxy.

\subsection{CMa is not the warp: GSC2.2 data}

The optical (photographic) photometry provided by the GSC2.2 catalogue is a much
less homogeneous dataset with respect to 2MASS. The effect of interstellar
extinction is much stronger in the GSC2.2 passbands ($J_{mag}$,$F_{mag}$) and
the corresponding reddening laws are not well known. 
On the other hand a direct test
performed on the F-XMM showed that the limiting magnitude of GSC2.2 photometry 
is $\sim 1.0$ mag fainter than that of 2MASS at $B-V\le 1.0$. 
Hence the inspection of
the GSC2.2 data may provide additional insight in the comparison between the CMa
overdensity and the Galactic warp. We have selected from the GSC2.2 two small 
circular regions of radius $r=1\degr$ toward the CMa and WARP directions and
their respective northern Control Fields. The fields have been selected to have
small and homogeneous reddening (see Tab.~2, for details). Comparison among the
overall LF of the four fields indicates that the completeness of the
samples is very similar down to $J_{mag}\le 17.5$. Small color shifts ($<0.2$
mag) have been applied to match the main features of the corresponding CMDs. We
stress that the result discussed below is completely independent of the
assumed shifts.

\begin{figure}
\includegraphics[width=84mm]{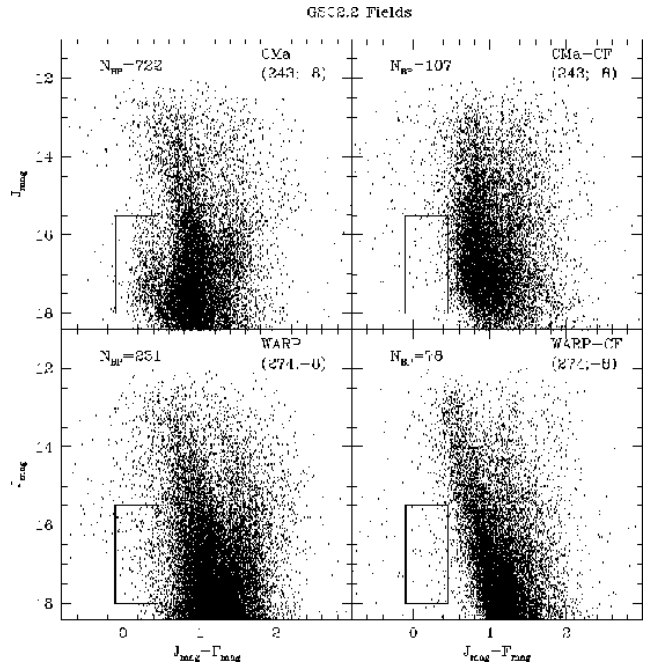} 
\caption{CMDs of the GSC2.2 selected fields described in Tab.~2. The overplotted
selection box encloses the Blue Plume feature. }
\end{figure}

The CMDs of the selected fields are shown in Fig.~14. A detailed analysis of the
diagrams suggests that there are differences in the stellar populations of the
various fields, particularly at faint magnitudes, and the study of LFs of
selected populations provides results fully compatible with those described in
the previous section, above. However, given the possible problems with reddening
corrections and the uncertain homogeneity of the samples\footnote{The same kind
of problems forced us to limit the comparisons with GSC2.2 data to smaller
fields with respect to the analysis of 2MASS data.}
we drop this part of
the analysis and concentrate on a single feature of the CMD, i.e. the Blue
Plume. In Pap-I and (above all) in Sect.~2 and 3.1 above we have shown that the
BP population is a distinctive feature of the CMa system. 
Fig.~14 shows that this
feature emerges clearly also from the GSC2.2 data: the well populated BP
observed in the CMa direction (upper left panel of Fig.~14) has no counterpart
in any of the other considered fields. In particular the excess of BP stars
(falling in the selection box reported in Fig.~14) in the CMa field with
respect to the WARP field is at the $\simeq 16 \sigma$ level, i.e. very large
and significant. This finding provides additional support to the conclusions of
the previous section and indicates that there is at least a significant
difference in the stellar populations hosted by the CMa system and by the
Galactic warp.

\subsection{Kinematics}

M04 derived proper motions of M giants in the CMa overdensity and found that the
tangential velocity of these stars is compatible with that of Galactic disc
stars at similar distance. Here we simply note that this observation does not
provide any conclusive argument on the nature of CMa since the uncertainties in
the measured proper motions leave room for differences in tangential velocity
as large as $\sim \pm 70$ km/s. 
In a dedicated contribution \citep{martinb} we show
that, on the other hand, CMa stars display a systemic radial velocity
significantly different from that of disc stars at the same distance, further
supporting the idea that CMa is an independent stellar system. 

\section{Clusters at the same distance as CMa}

In Pap-I it has been noted that there is also an apparent overdensity of open 
clusters in the surroundings of the CMa galaxy 
\cite[see also][]{frinch,forbes}. In the region of the sky
considered in the present analysis we found useful photometry only for two of
these clusters, namely Arp-Madore~2 (AM-2) and Tombaugh~2 (Tom~2). 

According to the most recent studies \citep{am2lee,am2} AM-2 is an old (age
$\simeq 5$ Gyr) and relatively metal poor ($\fe\simeq-0.5$) open cluster.
Its position in the sky [(l,b)=($248.2^o;-5.8^o$)] and its distance modulus 
\cite[$(m-M)_0=14.74\pm 0.16$][]{am2lee} place it fully within the CMa system,
according to the distance estimates derived in the present analysis. Note also
that the metallicity and age of this cluster are fully compatible with what 
found for the dominant population of the galaxy (see Sect.~2.5).
Unfortunately the field of view of the widest field photometry currently
available \citep{am2} is too small ($5.5^\prime \times 5.5^\prime$) to allow a
comparative study with the population surrounding the cluster (supposedly
mainly composed by CMa stars). The cluster population dominates the
field (in the range of magnitudes covered by the available CMD) at least out to
the edge of the field observed by \cite{am2} though the density profile of the
cluster appears to be essentially flat for $r> 1.5^\prime$ 
\cite[see, also][]{am2lee}.

The main characteristics of Tom~2 are quite similar to that of AM-2, but in this
case the wider field of view ($10.4^\prime \times 10.4^\prime$) of the best 
available photometry \citep{tom2} permits a more fruitful preliminary 
analysis, in the present context.
Tom~2 has the same metallicity as AM-2
\cite[$\fe=-0.5\pm 0.1$, from high resolution spectroscopy by][]{brown} and a
similar age ($\simeq 4$ Gyr, the quoted age difference with AM-2 is
probably consistent with zero within the uncertainties). It is located at 
(l,b)=($232.9^o;-6.9^o$) and its distance modulus 
\cite[$(m-M)_0=14.0\pm0.3$, according to][]{tom2} is consistent with that of the
CMa galaxy, within the errors. Also in this case the stellar population
is similar at any distance from the cluster center (within the available field)
though the profile is flat for $r\ga 160\arcsec$ \cite[see][]{tom2}.

In Fig.~15 the CMD of Tom~2 is compared with the CMD of F-XMM and with a
synthetic CMD from the R03 model of the Tom~2 field.
 The observed and synthetic CMDs of Tom~2 have been shifted to the same
reddening as F-XMM, correcting for the reddening differences indicated in the
plot.  The similarity of the Tom~2 population with that of
CMa is striking and suggests an intimate connection between the cluster and the
galaxy, confirming the hypotheses advanced by  \citet{martin} and 
\citet{frinch}.  

\begin{figure}
\includegraphics[width=84mm]{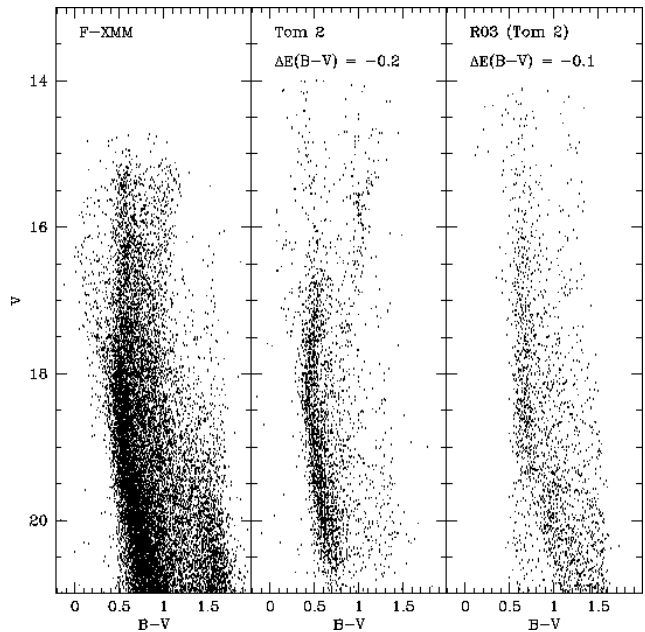} 
\caption{The CMD of the open cluster 
Tombaugh~2 (middle panel, from Kubiak et al., 1992) is compared with
the CMDs of the XMM filed (left panel) and of the synthetic CMD for
an equivalent field from R03. The
CMD of Tom~2 and the synthetic CMD have been shifted to the same reddening as
F-XMM by applying the reddening differences indicated in the panels.}
\end{figure}

All of the results and discussion presented above suggests that AM-2 and Tom~2 are
physically associated with the CMa system. A case strictly analogous to that of
Tom~2, i.e. a cluster having a population indistinguishable from the dominant
population of the parent galaxy, is that of Ter~7 in the Sgr dSph
\cite[see][]{marconi}. This cluster is also similar to AM-2 and Tom~2 in 
age and metal content. For Tom~2 the hypothesis may also be advanced that it
is not a genuine cluster but a mere overdensity within the Canis Major galaxy,
of the kind that have been observed in the Sgr dSph \citep{s2,sdgs1} and in the
UMi dSph \cite[see][and references therein]{umi}. The issue can be definitely
settled only with the observation of a significantly wider field surrounding the
cluster.

\section{Summary and Conclusions}

We have detected, for the first time, the Main Sequence of the main stellar 
population of the newly discovered galaxy in Canis Major (Pap-I), from
photometry of a wide field located at $\simeq 4.2\degr$ from its center
(the XMM field).
From the analysis of the CMD of the XMM field  we conclude that:

\begin{itemize}

\item The stellar population {\em observed} in the XMM field has no counterpart
      in the R03 Galactic model.

\item The distance modulus of CMa is $(m-M)_0\simeq 14.5\pm 0.3$,
      where the reported uncertainty  mainly reflects the uncertainty in the
      actual average metallicity of the system, that we assumed to lie
      in the range $-0.3\le [M/H]\le -0.7$, according to \citet{martin,crane}.
      The corresponding distance is $D_{\sun}\simeq 8.0 \pm 1.2$ kpc.
      
\item The age of the main population of CMa lies in the range 
      $4\sol$ age$\sol 10$ Gyr, slightly depending on the assumed metallicity.
      
\item There is a sparse population of younger stars (age$\sol 1-2$ Gyr) that is
      also associated with the CMa system (Blue Plume).

\end{itemize}

Tentative detections of stars related to the CMa/Ring system are presented for
the fields surrounding the open clusters NGC~2477, Tombaugh~1 and Berkeley~33.
The cases of two open clusters (AM~2 and Tombaugh~2) that may be physically 
associated with the CMa system have also been discussed. 
The results presented in this study provide independent confirmation, full
support and complement the results of \citet{martin}, that described
for the first time the CMa stellar system.

The claim by \citet{moma4} that the CMa overdensity could be entirely due to the
South/North asymmetry produced by the Galactic warp has been demonstrated to
lack the essential observational basis. Using the F-XMM data we have shown that
the warp model encapsulated in the R03 model (used by M04 to substantiate their
claim) fails to reproduce both the star counts {\em and} CMD morphology
observed in this field, located at $\simeq 4.2\degr$ from the center of the CMa
system. Using Red Clump and M giant stars from 2MASS we have shown that (a) the
CMa overdensity is much larger than that produced by the warp, and (b) it is 
found only at a well defined distance not compatible with the warp, fully
confirming the results of Pap-I and demonstrating that if all the available
information is taken into account the CMa overdensity clearly emerges {\em
above} the ``noise'' provided by the Galactic warp. Using GSC2.2 data we
demonstrated that the Blue Plume population is a distinctive characteristic of
the CMa system that has no comparable counterpart neither in Northern Control
Fields or in fields sampling the real warp.

A better characterization of the stellar content and star formation history of
the CMa system must wait for dedicated photometry and spectroscopy fully suited for
the study of this difficult target, immersed in very contaminated and
extincted regions of the sky. Here we tried to obtain the best possible
characterization of the system from existing and publicly available data, as
an immediate follow-up of the discovery.

Future searches for CMa stars in the background of other targets may prove
extremely useful 
in characterizing the spatial distribution of the galaxy and, in particular, of
the associated Ring. The relevant change of view is that what may be perceived
as an undesired contaminant of a given observation may hide more astrophysical
information than the target itself. This is true either if the background 
object
is an unknown stellar system (as CMa or Sgr) either if it is a poorly
characterized Galactic component, as the stellar warp of the Milky Way
\citep{binney,kuij}.
The lessons provided by the Sgr dSph (and
Stream) and by the CMa galaxy (and Ring) suggests to all of us a 
renewed attitude in the analysis of astronomical data, to avoid missing 
relevant pieces of information hidden in the background of our main targets.

\section*{Acknowledgments}

This work would not have been possible without the WEBDA database, developed and
maintained by J.-C. Mermilliod at the Geneva University, and without the effort
of many colleagues involved in the study of open clusters that made their data
publicly available through this database. 
We are grateful to A. Bragaglia and M. Tosi for many useful discussions
about open clusters and the Galactic Disk and to G. Carraro for providing
his photometry of Saurer's clusters in electronic form and in advance of
publication. Part of the data analysis has been performed using software
developed by P. Montegriffo at the Bologna Observatory (INAF).
The (anonymous) Referee is warmly thanked for comments and suggestions that
significantly improved the overall quality of the paper.

This research is partially supported by
the Italian {Ministero  dell'Universit\`a e della Ricerca Scientifica}
(MURST) through the COFIN grant p.  2002028935, assigned to the project {\em
Distance and  Stellar Populations in the galaxies of the Local Group}.
The support of ASI is also acknowledged.

This research has made use of ESO/EIS Pre-FLAMES data whose
observations have been carried out using the MPG/ESO 2.2m Telescope at the 
La Silla observatory under Program-ID No. 164.O-0561. 
This publication makes use of data products from the Two Micron All 
Sky Survey, which is a joint project of the University of Massachusetts and 
the Infrared Processing and Analysis Center/California Institute of 
Technology, funded by the National Aeronautics and Space Administration and 
the National Science Foundation.
The Guide Star Catalog was produced at the Space Telescope Science Institute 
under U.S. Government grant. These data are based on photographic data obtained 
using the Oschin Schmidt Telescope on Palomar Mountain and the UK Schmidt 
Telescope. 
The Oschin Schmidt Telescope is operated by the California Institute of 
Technology and Palomar Observatory. 
The UK Schmidt Telescope was operated by the Royal Observatory Edinburgh, 
with funding from the UK Science and Engineering Research Council (later the UK 
Particle Physics and Astronomy Research Council), until 1988 June, and 
thereafter by the Anglo-Australian Observatory. The blue plates of the southern 
Sky Atlas and its Equatorial Extension (together known as the SERC-J), as well 
as the Equatorial Red (ER) were all taken with the UK Schmidt. This research 
has made use of NASA's Astrophysics Data System Abstract Service. 

M.B. dedicates this work to Eugenio Pastore, who's walking with the
dogs of the Heavens.

\label{lastpage}

\end{document}